\pdfoutput=1
\documentclass[acs,a4paper,preprint,unsortedaddress,onecolumn]{revtex4-1}

\usepackage[fleqn]{amsmath}
\usepackage{amssymb}
\usepackage[dvips]{graphicx}
\usepackage{color}
\usepackage{tabularx}

\newcommand{\vect}[1]{\textbf{\textit{#1}}}

\newcommand{\period}[0]{T_{\textrm{P}}}

\newcommand{\confaa}[0]{{\alpha_{\textrm{R}}}}
\newcommand{\confab}[0]{{\alpha_{\textrm{R}}'}}
\newcommand{\confba}[0]{{\textrm{C}7_{\textrm{eq}}}}
\newcommand{\confbb}[0]{{\textrm{C}5}}
\newcommand{\confc}[0]{{\alpha_{\textrm{L}}}}

\begin{document}

\title{Exploring the conformational dynamics of alanine dipeptide in solution subjected to an external electric field: A nonequilibrium molecular dynamics simulation}
\author{Han Wang}
\affiliation{Institute for Mathematics, Freie Universit\"at Berlin, Germany}
\email{han.wang@fu-berlin.de}
\author{Christof Sch\"utte}
\affiliation{Institute for Mathematics, Freie Universit\"at Berlin, Germany}
\affiliation{Zuse Institute Berlin, Germany}
\author{Giovanni Ciccotti}
\affiliation{School of Physics, University College Dublin, Belfield, Dublin 4, Ireland}
\affiliation{Dipartimento di Fisica, Universit\`a ``La Sapienza'' and CNISM, Piazzale Aldo Moro 5, 00185 Roma, Italy}
\author{Luigi Delle Site}
\affiliation{Institute for Mathematics, Freie Universit\"at Berlin, Germany}
   
\begin{abstract}
In this paper, we investigate the conformational dynamics of alanine
   dipeptide under an external electric field by 
   nonequilibrium molecular dynamics simulation.  We consider
   the case of a constant and of an oscillatory field. In this context we
propose a procedure to
   implement the temperature control, which removes the irrelevant thermal
effects of the field. For the constant field different time-scales are
   identified  in the conformational, dipole moment, and orientational
   dynamics. Moreover, we prove that the solvent structure only
   marginally changes when the external field is switched on.  In the
   case of oscillatory field, the conformational changes are shown to be as
   strong as in the previous case, and non-trivial nonequilibrium
   circular paths in the conformation space are revealed by calculating the
   integrated net probability fluxes.
\end{abstract}

\maketitle

\section{Introduction}
The possible effect of the electric field (EF) on the conformation of
proteins is a subject of increasing attention since it is linked to the problem of exposing human tissues to
electromagnetic radiation. Despite its major relevance, a
satisfactory understanding of how the EF influences the behaviour of proteins
has not been reached so far although a considerable number of
experimental e.g.~\cite{bohr2000microwave, bohr2000microwave-1,
  dePomerai2000cell, inskip2001cellular, mancinelli2004non} and
theoretical e.g.~\cite{budi2005electric, budi2007effect,
  budi2008comparative, toschi2008effects, astrakas2011electric,
  astrakas2012structural, damm2012can, starzyk2013proteins,
  english2009nonequilibrium, solomentsev2012effects}
studies have been devoted to this topic.
In this perspective, molecular dynamics (MD) has been proven useful in
understanding the behaviour of a protein 
in both static and oscillatory EF~\cite{budi2005electric, budi2007effect, budi2008comparative,
  toschi2008effects, astrakas2011electric, astrakas2012structural,
  damm2012can, starzyk2013proteins, english2009nonequilibrium,
  solomentsev2012effects}. In these studies, the initial
configurations are usually taken from the Protein Data Bank (PDB),
and then temperature controlled simulations are performed for a maximum
time range of the order of 10-60 nanoseconds.
During these simulations time 
relevant conformational changes characterizing the solvated molecule can be observed. 
Along the trajectories, various observable, such as the
secondary structure, root mean square displacement (RMSD), dipole
moment, and shape parameters are calculated as a function of time, or
in stationary condition, taking time averages.
This approach releases useful information on how the
secondary structure of the molecule is changed by external  EF.  However,
the behaviour of observables in time is characterized by 
strong
fluctuations. The entity of the fluctuations is such that a quantitative 
analysis becomes difficult and does not allow for  a clear understanding of the rationale behind the conformational changes.
A time average can reduce the statistical uncertainty,
however in non-stationary conditions this would be a not-well-founded procedure.
In fact, we are studying non-stationary nonequilibrium process
(e.g. relaxing process to the new conformations under a static
EF, or the dynamics from one conformation to the other under an oscillatory EF).
In this perspective, if one wants to use MD, the proper way to do it would be that of treating a situation of nonequilibrium with algorithms that can explicitly describe it.
This is the approach that will be followed in this work, using
the dynamical nonequilibrium molecular dynamics simulation (D-NEMD).
The method, firstly developed by one of us and coworkers~\cite{ciccotti1975direct, ciccotti1979thought, ciccotti1993theoretical, palla2008bulk}, 
has been recently applied, to the study of
hydrodynamics~\cite{orlandini2011hydrodynamics,
  orlandini2011hydrodynamics-01}. The essential feature of D-NEMD is to
provide a way of evaluating time-dependent
nonequilibrium averages starting from a properly determined initial ensemble.
In this context, the method can be used to analyze the
conformational changes of a solvated molecule, subject to a time-dependent EF.

In the present paper, we study a small peptide, alanine dipeptide, as a model for showing how the method can be used. 
The choice of a small peptide, rather than a larger one, is done firstly because it
is easier to draw clear conclusions on the nonequilibrium properties
of a single peptide by excluding the slow-going interplay between
different peptide segments along a chain. 
Secondly, the computational cost required is rather massive and thus  testing
the various methodology proposed here would be hampered for large molecules,
although they remain our goal.  In any case, alanine dipeptide
is large enough so that the conformational changes of its internal degrees of freedom represent valid observables for a test of validity and utility of our approach.
The paper is organized as follows:
  We firstly describe the idea of D-NEMD, and propose a technical improvement consisting in the definition and application of a local thermostating process.
Next, after introducing the physical
observables chosen to analyze the nonequilibrium response,
the numerical results for the case of a static and oscillatory EF, respectively, are discussed. By calculating the nonequilibrium properties from D-NEMD
simulations, we are able to address
a few relevant questions: (i)~How the conformation of the peptide
changes in an EF (conformational space);
(ii)~When these changes happen (time scale),
and how likely they are (probability).
{One may wonder whether the information extracted from the D-NEMD simulation could have been
extracted just as well from conventional equilibrium simulations.
  In the appendix we show that in this case the D-NEMD approach shall be preferred to the equilibrium approach. The argument is provided by the comparison between the results of the nonequilibrium method and
  those from equilibrium simulations, within the linear response theory, based on the analysis of correlation functions.}
To our knowledge, this study is one of the few (if not the first)  where
time-dependent nonequilibrium w.r.t. conformation
is explicitly considered for a (relatively large) molecule
in solution subject to an external time-dependent EF.

\section{Methodology}
\subsection{Dynamical nonequilibrium molecular dynamics simulation (D-NEMD)}
In this section we briefly review the dynamical approach to nonequilibrium
molecular dynamics (D-NEMD) \cite{ciccotti1975direct, ciccotti1979thought,
  orlandini2011hydrodynamics, orlandini2011hydrodynamics-01}.
In the following we denote the macroscopic observable by $O(t)$. If at $t$
the configurational probability distribution is $\rho(\vect x, t)$, where
$\vect x$ is the phase space variable, then the observable can be
expressed by
\begin{align}\label{eqn:tmp1}
  O(t) = \int d\vect x\, \hat O(\vect x)\rho(\vect x, t)  = \langle \hat O(\vect x), \rho(\vect x, t)\rangle,
\end{align}
we refer to $\hat O (\vect x)$ as the microscopic observable, 
which is measured at each point in the phase space $\vect x$.
Here we always assume that the initial probability distribution
$\rho(\vect x, 0)$ is known. In particular, in our case,
it is identical to the equilibrium distribution of the system
without EF.
The bracket on the right hand side of Eq.~\eqref{eqn:tmp1} denotes, as usual, the inner product in the
phase space.  We assume that the dynamics of the system is governed by the
Hamiltonian equation, i.e. $\dot {\vect x} = J \cdot \nabla_{\vect x}
\mathcal H(\vect x, t)$, where $\mathcal H$ is the time-dependent Hamiltonian,
and $J$ is
the symplectic matrix
\begin{align}
  J = \left(
    \begin{array}{rr}
      0 & I\\
      -I & 0
    \end{array}
    \right).
\end{align}
The Liouville equation for the probability
distribution writes:
\begin{align}\label{eqn:tmp2}
  \frac{\partial \rho(\vect x, t)}{\partial t} = - iL(t) \rho(\vect x, t),
\end{align}
where $iL(t) = \{\cdot, \mathcal H\}$ is the Liouville operator.
Eq.~\eqref{eqn:tmp2}
can be formally solved by $\rho(\vect x, t) = U^\dagger(t,0) \rho(\vect x, 0)$,
{where $U^\dagger(t,0) = \mathcal T\exp\{-i \int_0^t d t' L(t')\}$, and $\mathcal T$ is the time ordering operator.}
On the other hand,
\begin{align}
  \frac{d \hat O(\vect x(t))}{dt} = \nabla_{\vect x}\hat O\cdot \dot{\vect x}
  = \nabla_{\vect x}\hat O\cdot J\cdot \nabla_{\vect x}\mathcal H
  = iL(t) \hat O (\vect x(t)).
\end{align}
This equation can be formally solved by
$\hat O(\vect x(t)) = U(t,0)\,\hat O(\vect x(0))$, therefore,
\begin{align}\nonumber
  O(t) & = \langle \hat O(\vect x), \rho(\vect x, t)\rangle
  = \langle \hat O(\vect x), U^\dagger(t,0)\, \rho(\vect x, 0)\rangle
  = \langle U(t,0)\,\hat O(\vect x), \rho(\vect x, 0)\rangle\\\label{eqn:tmp4}
  &= \langle \hat O(\vect x(t)), \rho(\vect x, 0)\rangle.
\end{align}
Since we assume that the system starts from the equilibrium distribution (without EF)
 Eq.~\eqref{eqn:tmp4} expresses the fact that the 
observable $O(t)$ calculated in a situation of nonequilibrium is equal to the ensemble average of
the microscopic observable computed along trajectories, starting from
initial configurations sampled from an initial or, sometimes, an \emph{equilibrium}
distribution. In practical terms we proceed by first running an equilibrium MD simulation
to generate a sample of configurations. Next we employ these configurations as
initial configurations and, for each, the full Hamiltonian dynamics is integrated until
time $t$.
In this paper, these trajectories are called \emph{branching
  trajectories}.
Finally, the value of the macroscopic observable at time $t$ is
calculated by averaging the microscopic observable measured at each time $t$ along each branching trajectory.
Stochastic dynamics (e.g.~Langevin dynamics) can be handled analogously.

\subsection{Control of temperature in nonequilibrium MD}\label{sec:tmp2b}

\begin{figure}
  \centering
  \includegraphics[width=0.3\textwidth]{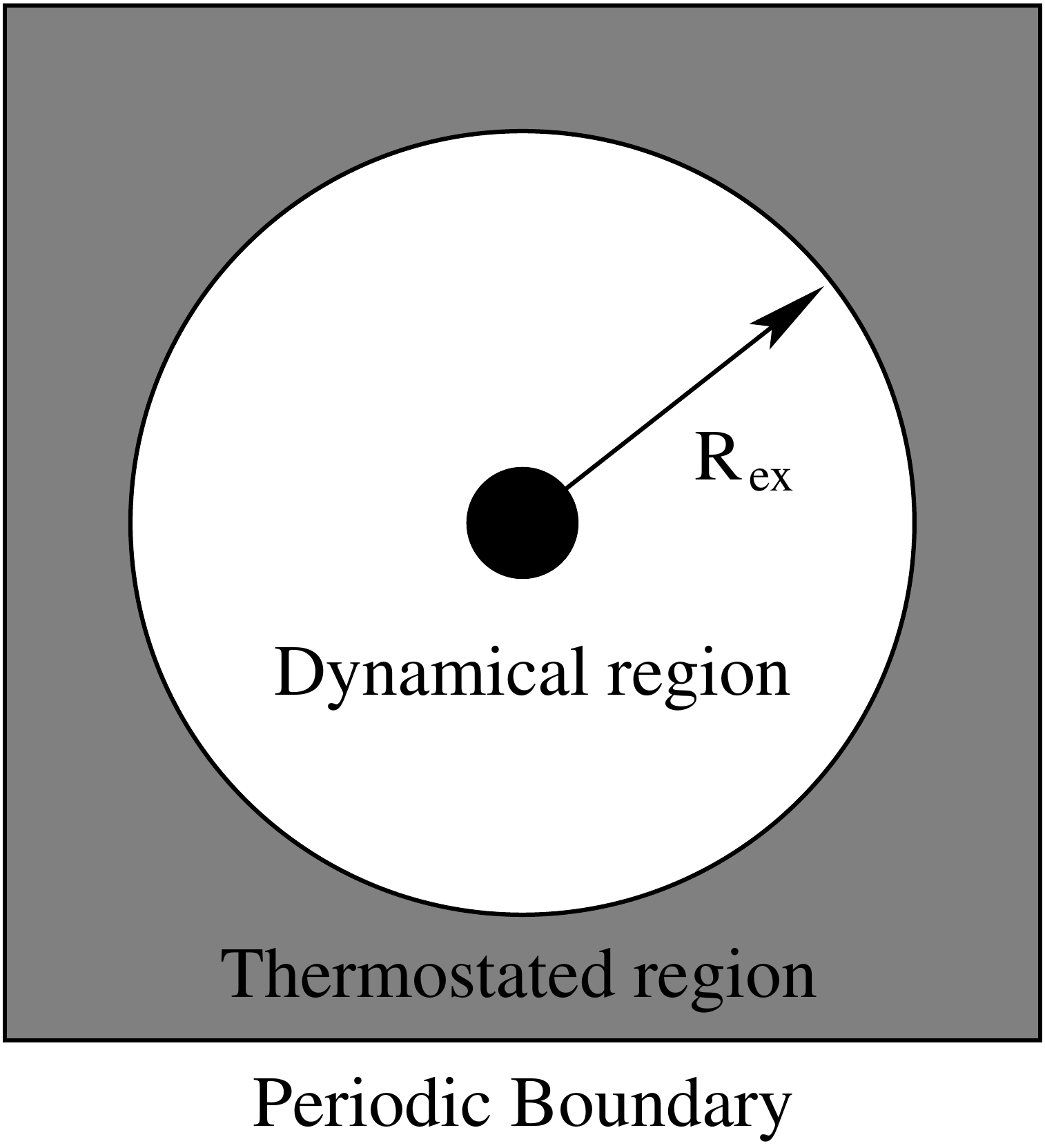}
  \caption{Schematic plot of the nonequilibrium temperature control scheme.
    The white area is the region characterized by pure dynamics without any direct action by the external thermostat.
    The gray area is the region where the thermostat is used to keep the target temperature.
    $R_{ex}$ denotes the size of the purely dynamical region, which
    is a sphere around the molecule of interest (in our case it is the alanine
    dipeptide).
    The whole system is subject to periodic boundary conditions. }
  \label{fig:tmp2}
\end{figure}
For this study, we are interested in 
nonthermal~\cite{delaHoz2005microwaves} effects of the EF
on the configuration of the solvated alanine dipeptide.
For this reason a proper control of the temperature in a nonequilibrium system
becomes an important issue; in fact it was argued that  an
accurate control of the internal reaction temperature is essential to
perform reproducible experiments~\cite{damm2012can}.
A straightforward (and realistic) implementation of the temperature control consists in 
embedding the alanine into an infinitely large solvent environment, and 
the EF is only applied to the neighborhood of the alanine. In this case
any extra heat generated by the EF can be effectively
dispersed in the large solvent environment.
In this way one can identify the effect of the EF on the conformational space of the solvated molecule,
without the artifact that the surrounding solvent cannot release the additional energy of the EF.
However, in practice, in most of the cases the computational cost of having large solvent environment is rather massive if not prohibitive.
Therefore, it is of interest to substitute the infinitely large
solvent environment with a smaller affordable system and at the same time accurately reproduce the nonequilibrium
physical properties of interest.
In MD, for equilibrium situations, the solution is to couple
the system to an external thermostat, under which the desired ensemble
(canonical in this case) can be
sampled by simulating a finite size system with periodic boundary
conditions.
However, coupling the system to a thermostat for nonequilibrium situations is a more delicate issue.
In fact, since a thermostat is explicitly designed to bring the system into
thermal equilibrium,
the perturbation produced by the EF would be mostly adsorbed by the thermostat. This means that the physics of the original dynamics in presence of the EF
would be lost and the response to the perturbation would be unphysical.

In this section we propose a procedure to solve the problem: we
start observing that the dynamics without thermostat of the alanine, 
and the water molecules in the closer solvation shells, is
crucial to determine the response of the observable of interest
under the EF perturbation
(in this work, the observable is the conformational change of alanine).
Instead the detailed dynamics of
the water molecules far away is not of primary importance: the region beyond the first solvation shells plays mainly the role of a thermodynamic bath. 
Therefore, instead of simulating an infinitely
large system, we simulate our finite size system with periodic
boundary condition, and with the following characteristics:
(i)~The simulating region is divided in two
subregions, a spherical region around the alanine dipeptide of radius $R_{ex}$,
centered at the alpha-carbon. Here the dynamics is
not subject to a thermostat.
We refer to such subregion as \emph{the  dynamical
region}. (ii)~Beyond $R_{ex}$, the dynamics of the water molecules is coupled
to a Langevin thermostat.  This region is called \emph{the thermostated
region}. For technical convenience we
fix the position of the alpha-carbon in space, so that the alanine is always
located at the center of the simulation region (see Fig.~\ref{fig:tmp2}).
In these conditions, the  dynamics is preserved in the dynamical region (where the properties of interest are observed), while 
possible artificial effects of the thermostating process in the outside region
are negligible due to the finite correlation
range of liquid water.
At the same time, the thermostated region works as a infinitely large
environment that effectively absorbs the extra heat produced by the EF.
The validity of the above
statements will be checked by a series of numerical tests where it is shown that
the response of observables in nonequilibrium (i.e. under the effect of the EF) does not depend on the size of the system
and on the size of the  dynamical subregion, provided that they are
reasonably large.
{ The proposed temperature control method has similarities with 
  the stochastic boundary condition proposed by Brooks and Karplus~\cite{brooks1983deformable}.
  However, differently from their approach, we do not explicitly consider a boundary region: The system is instead divided into a dynamical region (corresponding to the reaction region of \cite{brooks1983deformable}) and
  a thermostated region (corresponding to the stochastic buffer region in \cite{brooks1983deformable}) in a 3-D cubic periodic simulation box.
} 

\section{Case I: Alanine dipeptide
  under a uniform  constant EF}

\subsection{System settings and simulation protocol}
The system is set up in a $2.7\times 2.7\times
2.7\, \textsf{nm}^3$ periodic simulation region, with one alanine dipeptide
described by the CHARMM27 force field~\cite{foloppe2000all}, and dissolved in 644 TIP3P~\cite{jorgensen1983comparison}
water molecules. The grid-based energy correction map (CMAP)~\cite{mackerell2004extending}
for the backbone dihedral angles is also used.
The size of the  dynamical region is $R_{ex} = 1.0$~nm.
All simulations are performed by a home-modified GROMACS~4.5~\cite{pronk2013gromacs} together with the CHARMM
force field~\cite{bjelkmar2010implementation}.
First, an equilibrium NVT simulation at 300~K of
100~\textsf{ns} was performed with a Langevin thermostat (time-scale
$\tau_T = 0.5~\textsf{ps}$).  Along the trajectory, configurations were taken every
50~\textsf{ps} and we used 2000 initial configurations for each nonequilibrium
MD simulation (if not stated otherwise for specific cases).
The branching trajectories were integrated by the
Leap-frog scheme (standard Gromacs implementation) with the aforementioned nonequilibrium
temperature control technique.  The
time step was $0.002~\textsf{ps}$.
The short-range
interaction (van der Waals interaction)
have a cut-off radius of 1.0~nm, and has been smoothed from $0.8$ to $1.0$~nm by the
``switch'' method provided by the GROMACS code.
A energy conserving Particle Mesh Ewald (PME)~\cite{darden1993pme, essmann1995spm}
method was applied to calculate the electrostatic interaction in this
periodic system. For the direct space part of PME the cut-off and smoothing follow the same principles as those applied to the van der Waals interaction.
All hydrogen involving bonds are constrained by LINCS~\cite{hess1997lincs}, except 
the TIP3P water molecules which are constrained by the SETTLE algorithm~\cite{miyamoto2004settle}.
In the thermostated region, the original dynamics was
coupled to a Langevin thermostat with $\tau_T = 0.1~\textsf{ps}$.
In all testing cases, this local thermostat is able to control
the system at the desired temperature, i.e.~300~K (the results
are not presented in this paper).
The whole system is also coupled to a Parrinello-Rahman barostat~\cite{parrinello1981polymorphic}~(in standard Gromacs implementation) with $\tau_P = 2.0~\textsf{ps}$ to keep
the pressure at ambient condition (1~Bar). Since the
change of the system size is small and slow, the pressure control
does not have an sizable effect on the
dynamics of the system.
At time $t=0$~ps, the system
has been fully equilibrated without any EF. From $t=0$ to
the warm-up time $t=t_{\textrm{warm}}$, the EF is switched on linearly, while
after $t=t_{\textrm{warm}}$, the field is kept constant in time at
$E_{\infty}$.  In this work we consider $t_{\textrm{warm}} = 10$~ps,
and $E_{\infty} = 1$~V/nm.
The direction of the field is arbitrarily chosen along the
$x$ direction.
In this paper we denote the
EF as a function of time $\vect E(t)$.
Therefore, for the case of constant EF, we have
$\vect E(t) = (E_\infty\cdot t/t_{warm},0,0)$ for $0\leq t < t_{warm}$, and 
$\vect E(t) = (E_\infty,0,0)$
for $t \geq t_{warm}$.

\subsection{Molecular conformation and net probability flux}

\begin{figure}
  \centering
  \includegraphics[width=1.00\textwidth]{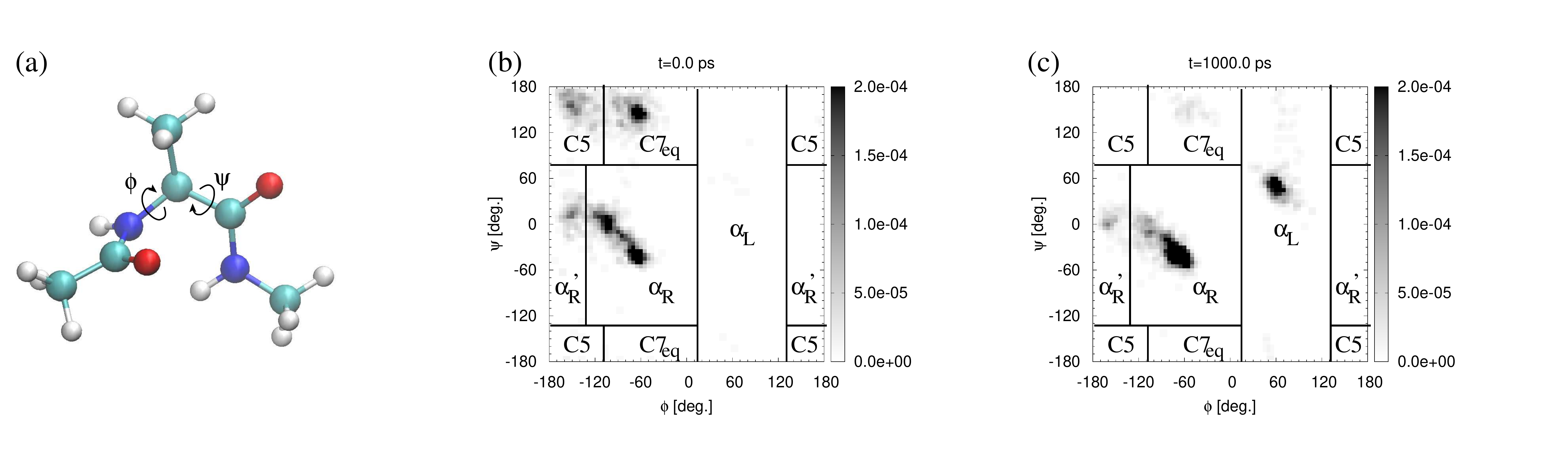}
  \caption{
    The probability density of the conformations of the alanine dipeptide plotted on the Ramachandran plot.
    (a): Two dihedral angles
    $\phi$ and $\psi$ are used to represent the molecular conformations of the alanine dipeptide.
    (b): The equilibrium probability density 
    at $t=0$~\textsf{ps}. (c): The probability density of equilibrium at $t=1000$~\textsf{ps}, resulting from the application of a constant EF of 1~V/nm. Darker color
    indicates higher probability.
    Plots are divided into 5 subregions corresponding
    to 5 molecular conformations:
    $\confaa = \{(\phi, \psi) | -134^\circ \leq \phi <  13^\circ, -125^\circ \leq \psi < 74^\circ\}$,
    $\confab = \{(\phi, \psi) |  128^\circ \leq \phi < 226^\circ, -125^\circ \leq \psi < 74^\circ\}$,
    $\confba = \{(\phi, \psi) | -110^\circ \leq \phi <  13^\circ,   74^\circ \leq \psi < 235^\circ\}$,
    $\confbb = \{(\phi, \psi) |  128^\circ \leq \phi < 250^\circ,   74^\circ \leq \psi < 235^\circ\}$,
    $\confc   = \{(\phi, \psi) |   13^\circ \leq \phi < 128^\circ, -180^\circ \leq \psi < 180^\circ\}$.
    Considering the angular periodicity, the angle, for example, $226^\circ$ corresponds to $-134^\circ$
    on the plot (b) and (c).
    This division is the same for both plot (b) and (c).
  }
  \label{fig:tmp4}
\end{figure}

The change of conformation in time of the alanine dipeptide is
investigated by analyzing the probability density $p(\phi,\psi,t)$ on
the Ramachandran plot at time $t$. The definition of the dihedral
angles $\phi$ and $\psi$ is given in Fig.~\ref{fig:tmp4}~(a).
The probability density of conformations of equilibrium in absence of external field is shown in Fig.~\ref{fig:tmp4}~(b), while the same quantity, resulting from the molecular relaxation to the action of $E_{\infty} = 1$~V/nm is
given in Fig.~\ref{fig:tmp4}~(c). In the plot one can identify several clusters
in which conformations are grouped.
Therefore, in order to simplify the analysis, we divide the
Ramachandran plot into 5 subregions $\{\confaa, \confab, \confba, \confbb, \confc\}$
(see caption of Fig.~\ref{fig:tmp4} for the corresponding definition).
It is worth to notice that these conformations
are found in different secondary structures of a peptide chain:
$\confaa$ and $\confab$ 
correspond to the $\alpha$-helix. $\confba$ and $\confbb$ 
correspond to the $\beta$-sheet. $\confc$ corresponds to the
left-handed $\alpha$-helix.

From Fig.~\ref{fig:tmp4}, it is evident that the probability with which a
conformation occurs changes when the external EF is applied.
The most evident case is conformation $\confc$; in fact it is {almost} not present before the external field is applied but appears in a clear way as a response to the action of the EF. Following this line of thought, we are interested to study the change of conformation under the action of EF. Specifically, for each conformational change, we will analyze the relation between the structural relaxation of the molecule and its corresponding time scale. 
The starting point is the calculation of the probability
of each conformation:
\begin{align}
  P_I(t) = \mathbb P ( (\phi_t,\psi_t) \in I), \quad  I \textrm{ being one of the five regions } \{\confaa, \confab, \confba, \confbb, \confc\},
\end{align}
Notice that the probability is time dependent, being an observable in nonequilibrium situation.
Next,
we consider the \emph{net probability flux} from conformation $J$ to $I$, defined by:
\begin{align}\nonumber
  F_{J,I}(t) & = \frac1{\Delta t}\,
  \bigg[\,
  \mathbb P \big( (\phi_{t-\Delta t},\psi_{t-\Delta t}) \in J,\, (\phi_t, \psi_t) \in I \,\big) 
  -
  \mathbb P \big( (\phi_{t-\Delta t}, \psi_{t-\Delta t}) \in I,\, (\phi_t, \psi_t) \in J \,\big)
  \bigg], \\\label{eqn:tmp7}
  & J,I \in \{\confaa, \confab, \confba, \confbb, \confc\}.
\end{align}
Where $\mathbb P\big( (\phi_{t-\Delta t}, \psi_{t-\Delta t}) \in J,\, (\phi_t, \psi_t) \in I\,\big)$
is the joint probability of the alanine being in conformation $J$ at
time $(t-\Delta t)$ \emph{and} being in conformation $I$ at time $t$.
The same definition applies to
$\mathbb P \big( (\phi_{t-\Delta t}, \psi_{t-\Delta t}) \in I,\, (\phi_t, \psi_t) \in J\,\big)$.
Here the time interval $\Delta t$ should be small enough compared to the
time scale of conformational dynamics,
so that the changes in the conformation probabilities
can be treated in linear approximation.
At the same time,  $\Delta t$ should be also large enough compared with
the time step of the MD integrator, 
so that the quantities of Eq.~\eqref{eqn:tmp7} are well-defined, and can be
estimated with sufficient numerical accuracy (see also Ref.~\cite{schuette2011markov}).
{In all the numerical examples of this work, $\Delta t = 1$~ps is used}.
A positive value of $F_{J,I}(t)$ indicates that the
net flux goes from $J$ to $I$,
while a negative value indicates a net flux from $I$ to $J$.
Through $F_{J,I}(t)$, the analysis of the nonequilibrium process 
is projected onto the analysis of the probabilistic link between discretized conformations $\{\confaa, \confab, \confba, \confbb, \confc\}$.
The concept of net probability flux employed by us is very close to the
  concept of kinetic rate.  However, we prefer to stick to our definition which
is mathematically simpler and univocal.
In the case of oscillatory EF, the behaviour of the net probability flux is also highly oscillatory, thus, in order to clearly identify the essential features of the molecular conformational changes, it is convenient to study the integrated net probability flux, defined as:
\begin{align}\label{eqn:tmp7a}
  Q_{J,I} (t) = \int_0^t F_{J,I}(\tau)d \tau,
\end{align}
which expresses the cumulative effects due to the action of EF on the conformations of the molecule.

\subsection{Conformational dynamics}

\begin{figure}
  \centering
  \includegraphics[]{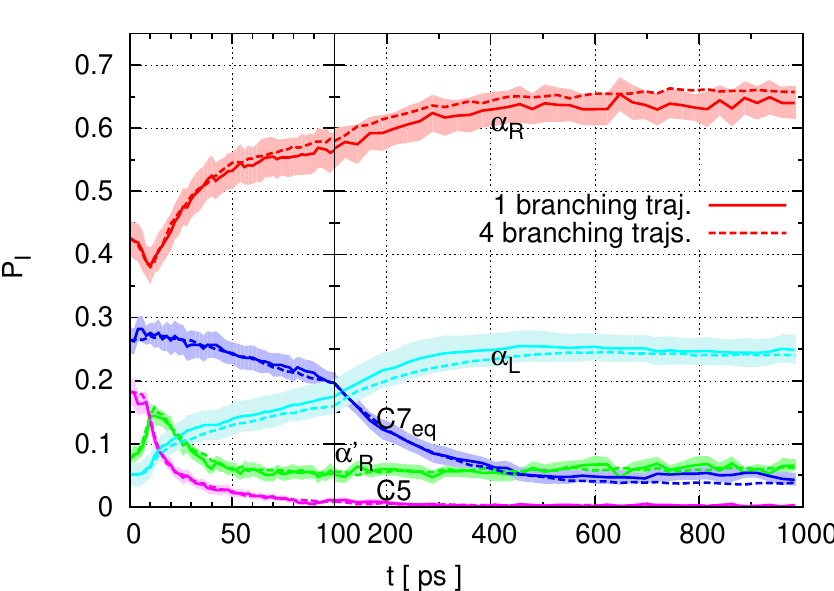}
  \caption{The time-dependent probability of conformations $P_I(t)$ 
    when a constant EF is present, where $\sum_I P_I(t) = 1, \forall t$.
    The warm-up time is $t_{warm} = 10$~ps. The red line refers to conformation $I = \confaa$,
    green to $\confab$, dark blue to $\confba$, purple to $\confbb$ and light blue
    to $\confc$. {
      The solid lines denote the results of one branching trajectory
      starting from each initial conformation, while the dashed lines denote
      the average of the four branching trajectories (different random seeds for the Langevin thermostat, see text) starting from each initial
      conformation.
      The shadowed region with each solid line denotes the
      statistical uncertainty of the result at 95\% confidence level.
    }
  }
  \label{fig:tmp5}
\end{figure}

\begin{figure}
  \centering
  \includegraphics[width=0.3\textwidth]{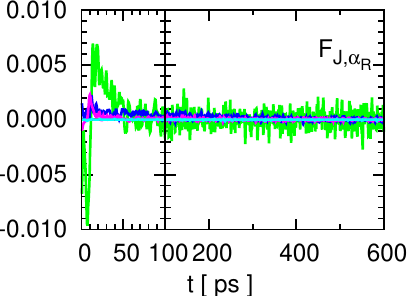}
  \includegraphics[width=0.3\textwidth]{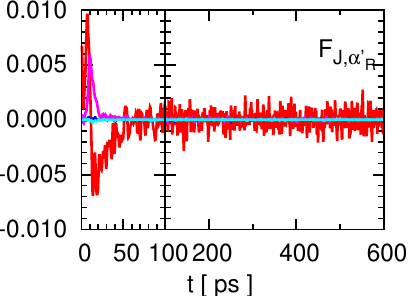}
  \includegraphics[width=0.3\textwidth]{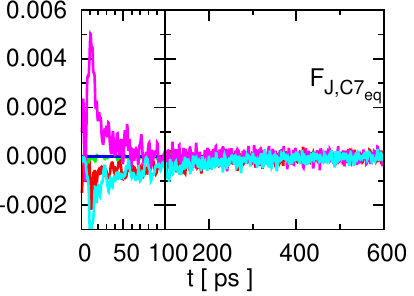}
  \includegraphics[width=0.3\textwidth]{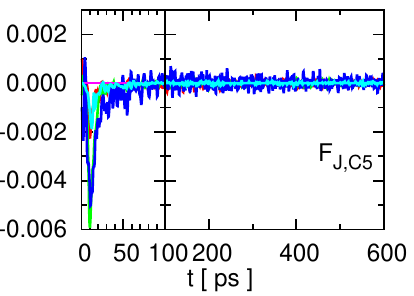}
  \includegraphics[width=0.3\textwidth]{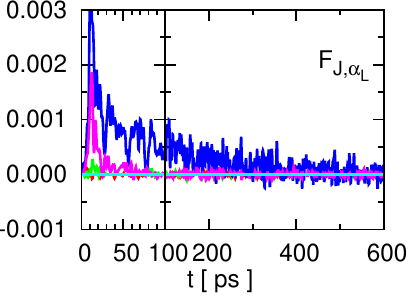}
  \caption{
    The net probability flux for the constant EF case.
    The warm-up time is $t_{warm} = 10$~ps.
    The probability flux $F_{J,I}(t)$ (defined by Eqn.~\eqref{eqn:tmp7}) is reported in units of $\textrm{ps}^{-1}$.
    From top, left to right we report $I = \confaa$, $\confab$, $\confba$; bottom, left to right $\confbb$ and
    $\confc$, respectively. In each plot, the red line stands for $J=\confaa$,
    green for $J=\confab$, dark blue for $J=\confba$, purple for $J=\confbb$ and light blue
    for $J=\confc$.
    {This plot is drawn from the four-branching-trajectories case (see the text).}
    }
  \label{fig:tmp6}
\end{figure}
The time-dependent probability density of the system to stay in a certain
conformation is given in
Fig.~\ref{fig:tmp5}.
In order to show that  our simulation results are robust with respect to the randomness introduced by the (local) Langevin thermostat, we have tested the effects of changing random seeds for the thermostat in the thermostated region. We compared simulation results obtained by starting a branching trajectory
from each initial conformation with those obtained from four different branching trajectories from each initial conformation (each of four with
different random seed for the Langevin thermostat in the thermostated region).
Results are consistent within the statistical uncertainty (see Fig.\ref{fig:tmp5}).
The $\beta$-sheet conformations $\confba$ and $\confbb$,
which are present at the initial stage after EF is switched on,
{fade away as the system relaxes}. The
probability of the $\alpha$-helix conformation $\confaa$, grows from 0.45 to 0.69.
The $\alpha$-helix $\confab$  does not change significantly  under the action of EF.
The left-hand helix conformation $\confc$ noticeably grows from 0 to 0.22.
We plot the net probability fluxes {of the four-branching-trajectories case} in Fig.~\ref{fig:tmp6}.
Results show that for short $t$, the net fluxes are generally non-zero,
however, 
as $t$ goes to infinity,
all fluxes converge to zero. This indicates
that as the EF is switched on, the system is driven away from the initial
equilibrium (at zero EF),  starting a dynamical nonequilibrium process.
After sufficiently long time, the system
is fully relaxed to the new stationary state, in
which it remains as long as the EF is switched on.
Of course we cannot exclude very slow nonequilibrium processes,
which cannot be captured by the duration of
our nonequilibrium simulation (1~ns), since it is
obviously that the proposed nonequilibrium approach
can only study time-dependent behaviours that are shorter than
the total time of the branching trajectories. 
An important point is that despite the system is finally relaxed to the new conformation,
the relaxation timescales corresponding to each conformation are rather different. 
From Fig.~\ref{fig:tmp5} and \ref{fig:tmp6}, we observe mainly
three different timescales 10~ps, 100~ps and 500~ps,
whose corresponding conformations are summarized in Tab.~\ref{tab:tmp1}.
This is a relevant result because it shows the possibility of employing an external EF as a tool to identify relevant time scales in the conformational behaviour of a molecule in solution.

\begin{table}
  \centering
  \begin{tabular*}{0.4\textwidth}{@{\extracolsep{\fill}}lc}\hline\hline
    Direction        & time scale [ ps ] \\\hline
    $\confaa\rightarrow \confab$        &       $\sim 10$      \\
    $\confbb\rightarrow \confaa,\ \confab,\ \confc$        &       $\sim 50$      \\    
    $\confab\rightarrow \confaa$        &       $\sim 100$      \\
    $\confbb\rightarrow \confba$        &       $\sim 100$      \\    
    $\confba\rightarrow \confaa,\ \confc$        &       $\sim 500$      \\    \hline\hline
  \end{tabular*}
  \caption{A list of the main probability fluxes and the corresponding time scales observed in the constant EF case.}
  \label{tab:tmp1}
\end{table}

\subsection{Molecular dipole response}
The fact that our external perturbation corresponds to the action of
an electric field, naturally leads to the question 
  of the alignment of the molecular dipole vector along the direction of external EF, and, in turn, of how this is related
to the overall conformational change of the molecule as reported in
the previous section.
\begin{figure}
  \centering
  \includegraphics[width=0.45\textwidth]{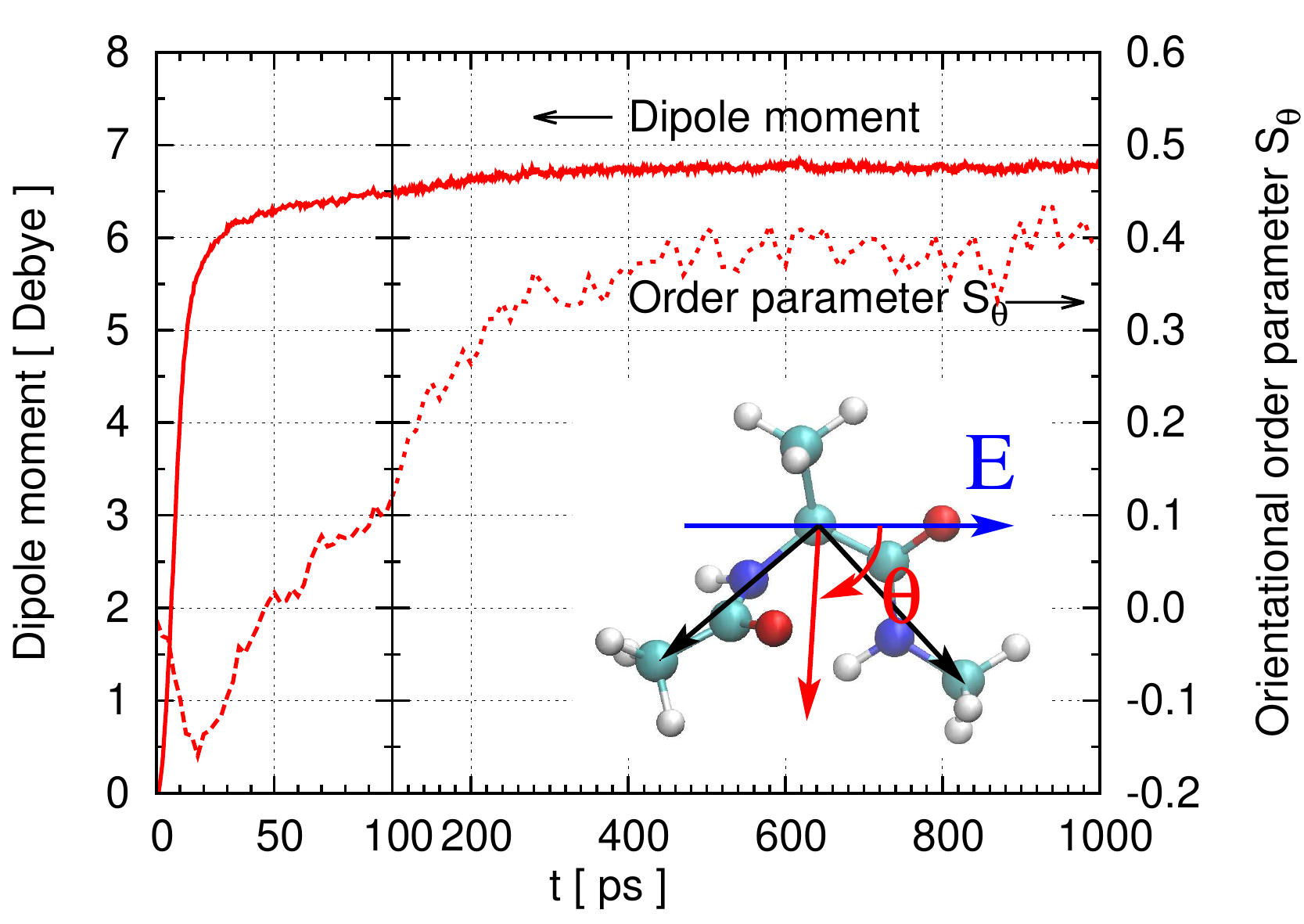}
  \caption{The
    $x$-component of the molecular dipole moment, and the orientational order parameter
    of the alanine molecule as a
    function of time. The solid line represents the dipole moment,
    while the dashed line represents the orientational order parameter.
    The left vertical axis is for the dipole moment, while
    the right is for the orientational order parameter.
  }
  \label{fig:tmp3}
\end{figure}

The dipole moment of the alanine dipeptide is defined by
\begin{align}\label{eqn:tmp8}
  \boldsymbol\mu_{\textrm{alanine}}(t) =
  \bigg\langle
  \sum_{i\in \{\textrm{point charges of alanine}\}}
  q_i\, \vect r_i(t)
  \bigg\rangle,
\end{align}
where $q_i$ denotes the partial charge of any point charge $i$ defining our model of alanine,
the molecule being neutral, i.e.~$\sum_iq_i = 0$.
$\vect r_i$ denotes the position of the point $i$.
Fig.~\ref{fig:tmp3}
shows that the $x$-component of the dipole moment 
reaches 85\% of its maximum value in only 20~ps, which is comparable to
the warm-up time $t_{warm}$, and is 25 times
smaller than the slowest time scale of the conformational relaxation.
Then in the following 400~ps, the dipole slowly relaxes to its 
maximum value, that is, 
6.8~Debye, and  corresponds to an energy
($\boldsymbol\mu_{\textrm{alanine}} \cdot \vect E$)
of c.a.~$-13.6$~kJ/mol.
We calculated the averaged dipole moment of different conformations
(defined by taking averages in Eqn.~\eqref{eqn:tmp8}
only for given conformations)
under constant EF, and found 6.8, 6.0, 5.1, 3.1 and 7.1~Debye
for $\confaa$, $\confab$, $\confba$, $\confbb$ and $\confc$, respectively.
The dipole energy difference between 3.1 and 7.1~Debye is roughly 8.0~kJ/mol.
Under a constant EF, the system will be likely to be driven towards those
conformations with higher dipole moment, because the
energy of the system will be lowered.  We suggest that this energy difference may be the reason why
we observe that the $\beta$-sheet conformations with lower dipole moment
are driven towards $\alpha$-helix conformations
$\confaa$ and $\confc$, whose dipole moments are the highest
among all conformations. Moreover, the $\confbb$ conformation vanishes because
its dipole is significantly lower than other conformations.
Comparing Fig.~\ref{fig:tmp3} with~\ref{fig:tmp5},
  the sharp increment of dipole moment before 20~ps is due to the quick vanishing of conformation
  $\confbb$, while the slow increment of dipole moment until 400~ps is due to
  the slow migration from conformation $\confba$ to $\confaa$ and $\confc$.

\subsection{Response of the orientational order parameter}
As a complementary information to the behaviour of the molecular dipole moment it is of interest to describe the overall orientational behaviour of the molecules w.r.t. the EF. In fact while the dipole moment specifically expresses the rearrangement of charges within the molecule as a response to the EF, the overall direction of the molecule tells us about the interplay between the internal positional rearrangement of the atoms and their alignment w.r.t. the EF.   
To this aim, let us define
the geometric direction of the alanine by the red vector in the inset in
Fig.~\ref{fig:tmp3}, i.e.~by the angle bisector of the two black vectors,
which connect the $\alpha$-carbon and the carbons
on the methyl groups. 
We considered the angle $\theta$ made by 
the orientation of the alanine dipeptide in space w.r.t.~the direction of EF
as a function of time.
Then we defined an orientational order parameter 
\begin{align}
  S_\theta = \langle 3\sin^2\theta - 2\rangle.
\end{align}
The ensemble average is made along the branching trajectories, so it is
a time dependent observable.
The order parameter
indicates the (time dependent) average orientation of the molecule w.r.t. ~the direction of the EF.
If the molecule is perfectly perpendicular to the EF, then $S_\theta = 1$, if is perfectly parallel then $S_\theta = -2$, if it has no directional
preference, then $S_\theta = 0$. 
From $t=0$ to roughly 10~ps, the orientational order parameter rapidly
decreases from 0 to $-0.15$, which means a weak alignment of the molecule to the external field.
Then from $t=10$ to 500~ps, the molecule slowly changes to the orientation
that is almost perpendicular to the EF.
In fact,
  the vector of the geometric direction tends to be perpendicular
    to the dipole moment,
  if one observes the correlation between the value
  of the dipole moment (which tends to be more parallel to the EF as
  its value increases) and the angle of the vector of the geometric
  direction, the result above can be easily explained.
This orientational change is linked to the observed slowest
time scale of the molecular conformational change.

\subsection{Effects of EF on the solvent around the molecule}

\begin{figure}
  \centering
  \includegraphics[width=0.48\textwidth]{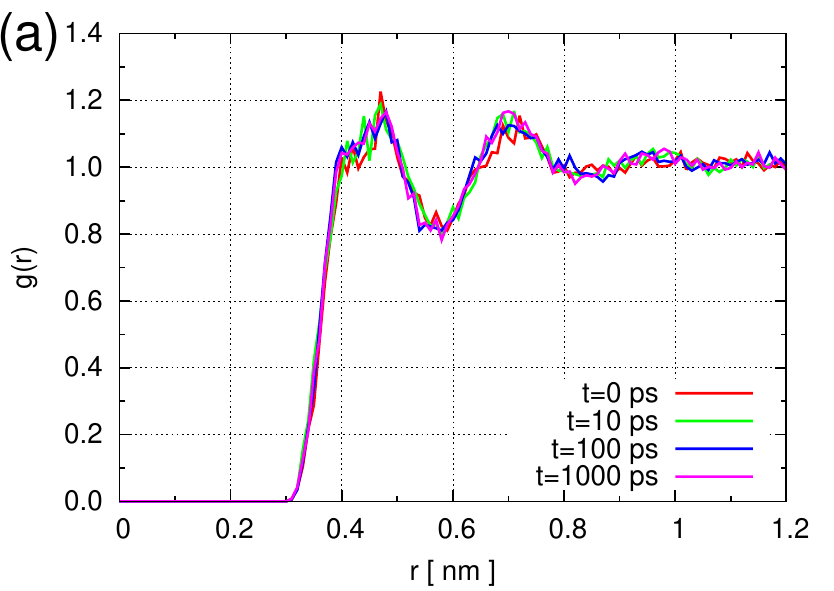}
  \includegraphics[width=0.48\textwidth]{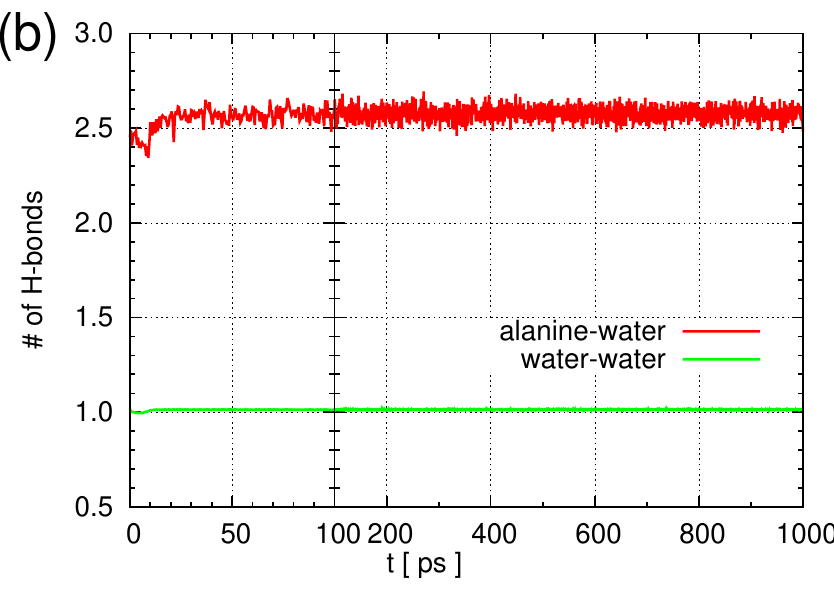}
  \caption{The solvent structure around the alanine dipeptide.  
    (a): the radial distribution function (RDF) between the
    $\alpha$-carbon and the center-of-mass (COM) of the water molecules at
    $t = 0$, 10, 100 and 1000~ps.  (b): the number of hydrogen
    bonds formed between the alanine and water, and between two water
    molecules.  The number of hydrogen bonds between two water
    molecules is normalized by the number of water molecules in the
    system. 
  }
  \label{fig:tmp7a}
\end{figure}
So far we have focused our attention on the behaviour of the alanine
under the action of the EF. However, it is also important to understand
the effects of the EF on the solvating water molecules. Being water a
polar molecule the electrostatic interaction with an EF can
dramatically change its solvation structure around the alanine and
thus, in turn, influence its conformational behaviour. For this
reason, in this section we analyze the behaviour of the solvent around
the alanine under the action of the EF by investigating
the radial distribution function (RDF) between $\alpha$-carbon and the
center-of-mass (COM) of the water molecules
(see Fig.~\ref{fig:tmp7a} (a)). The average
RDF does not change significantly as a function of time
(within the statistical error). 
Fig.~\ref{fig:tmp7a}~(b) presents the number of hydrogen bonds between the
alanine and water molecules, and between two neighboring water molecules.
The number of hydrogen bonds are calculated by standard GROMACS
routine {\texttt {g\_hbond}}, with donor-acceptor distance cut-off 0.3~nm,
and hydrogen-donor-acceptor angle cut-off $20^{\circ}$.
We observe that the
number of hydrogen bonds remains almost the same when the EF is turned on,
although molecular dipoles tend to align along the direction of the EF (result
not shown). 
The EF-induced solvent effect on the alanine can be also accounted by calculating the
difference of the water-alanine electrostatic interaction 
without and with the EF.
The energy difference is found to be $-19.1$~kJ/mol that is
comparable to the electrostatic energy associated to the dipole of the alanine itself.
Therefore, we conclude that the solvation structure around the
alanine does not change or contribute significantly in determining the conformational dynamics of the alanine under the action of EF.

\subsection{The finite-size effect on the dynamical and thermostated regions}

\begin{figure}
  \centering
  \includegraphics[]{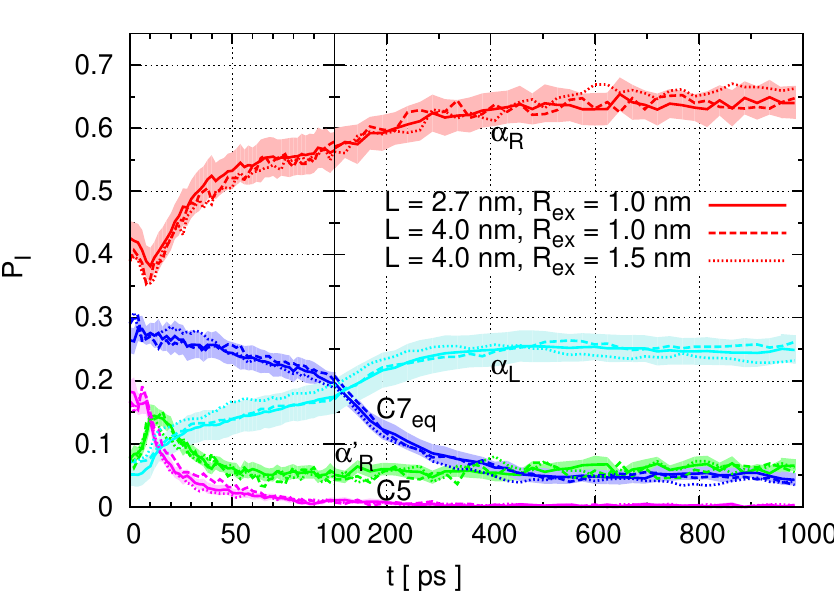}
  \caption{Testing the finite-size effect on the dynamical and thermostated region.
    Probabilities of different conformations are presented versus time.
    See the text for more details.
    {The shadowed region denotes the
      statistical uncertainty of the $L=2.7$~nm $R_{ex}=1.0$~nm case at 95\% confidence level.
    }
  }
  \label{fig:tmp7}
\end{figure}

Just as described in Sec.~\ref{sec:tmp2b}, we performed nonequilibrium MD
simulations in a finite size periodic system, further divided 
into a dynamical and a thermostated region.
Since the periodic boundary conditions and the division of the system
are technical approximations to the real system with dynamics-preserving thermal control, 
the effect of the finite-size in 
the system settings should be carefully checked.
Therefore, we
perform two additional simulations: one with box dimensions
$L=4.0$~nm (the box is cubic) and a
Hamiltonian dynamical region of radius $R_{ex} = 1.0$~nm. A second simulation is done with box size $L=4.0$~nm and a dynamical region of
radius $R_{ex} = 1.5$~nm. We compare the results of these two simulations to those of the system we have used so far, i.e. $L=2.7$~nm and $R_{ex} = 1.0$~nm.
Fig.~\ref{fig:tmp7} shows essential consistency in
the simulation results for the three systems. The finite-size effects are
negligible.

\section{Case II:
  periodically oscillatory EF}
A step forward in our study is to consider a periodically oscillatory EF which has
a $\sin$-wave shape:
\begin{align}
  \vect E(t) = (E_0\sin(2\pi t / \period), 0, 0),
\end{align}
where $E_0$ is the intensity of the field, which is chosen to be
1.0~V/nm.  $\period$ is the oscillating period.
Here we tested three different
periods 10, 40 and 200~ps.
\begin{figure}
  \centering
  \includegraphics[width=0.48\textwidth]{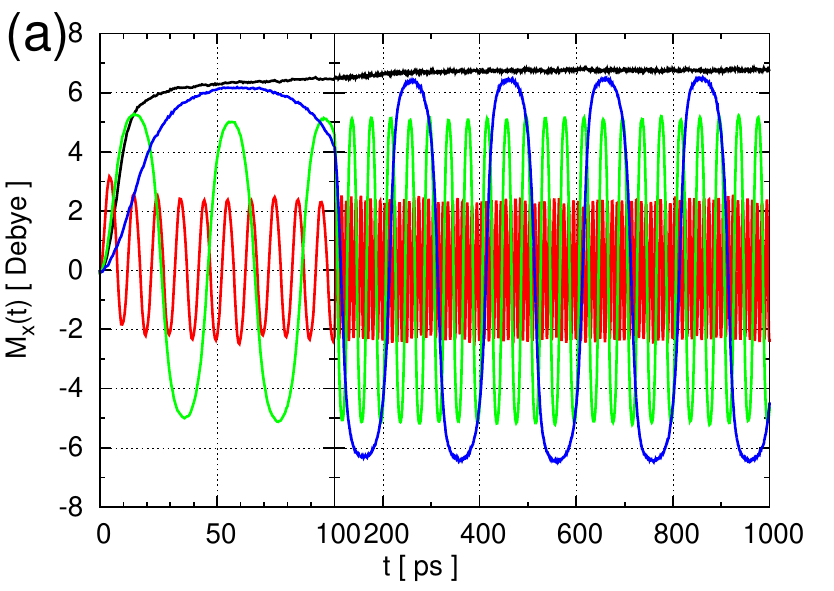}
  \includegraphics[width=0.48\textwidth]{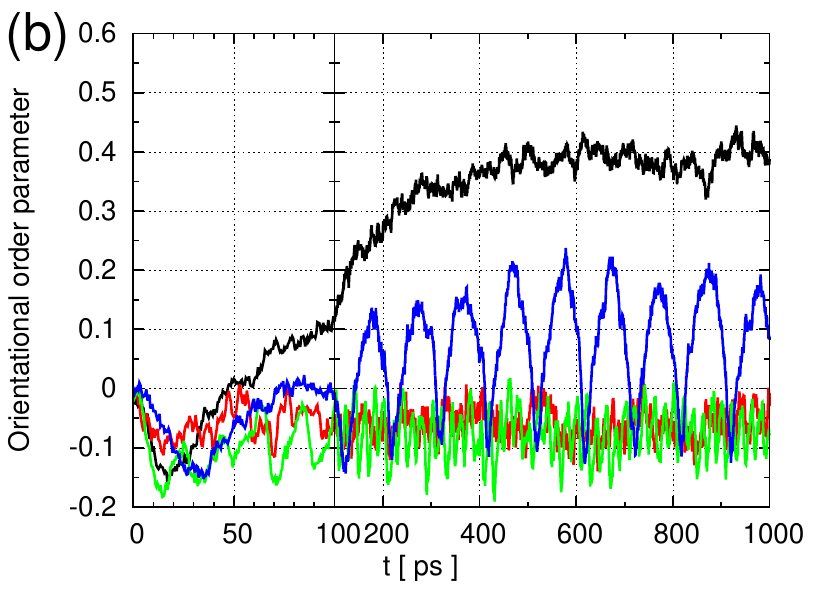}
  \caption{Illustration of the effect of the oscillatory EF on the system:
    (a) the dipole moment  and (b) the
    orientational order parameter of alanine dipeptide as a
    function of time. In (a),
    only the $x$-component of the dipole moment is
    shown. The black line refers to the static EF which is turned on
    at time $t_{warm} = 10$~ps. The red line refers to the oscillatory field with period of
    10~ps. The green line is of 40~ps; and the blue of 200~ps.}
  \label{fig:tmp8}
\end{figure}
Fig.~\ref{fig:tmp8} (a)
presents the $x$-component of the molecular dipole moment as
a function of time. The red, green and blue lines report the
results of $\period =10$~ps, 40~ps and 200~ps, respectively. The black line
shows the dipole moment under constant EF
for reference.
It shows that the oscillating period of the molecular dipole moment
is the same as the period of the oscillatory EF. Therefore,
the molecular dipole moment
is able to respond to the external EF almost immediately.
At $\period =200$~ps, the maximum magnitude of the molecular
dipole moment is close to the value obtained in the case of constant EF, which
means that the variation of the oscillatory EF is so slow that the
alanine has enough time to 
relax its dipole. However,
for $\period =10$ and 40~ps, the EF oscillates faster and the molecule 
does not have enough time to fully relax the dipole, as a consequence the maximum dipole moment is smaller than the previous case.
Fig.~\ref{fig:tmp8}~(b) presents the orientational order
parameter. The notations are the same as Fig~\ref{fig:tmp8}~(a).
For all cases the order parameter is much
smaller than in the case of constant EF. One possible reason is that the
relaxation of the order parameter is very slow, and the molecule in the case of oscillatory field is
not exposed to a strong enough EF for a time long enough.
For  $\period =10$ and 40~ps, the orientation of the alanine shows a {very} weak tendency
to be parallel to the EF. It is worth to notice that for constant EF,
we also observe a quick alignment of the molecular orientation vector to the EF
on the time scale of 10~ps. For $\period =200$~ps, the molecule tends periodically
to be more perpendicular to the EF, but this directional preference is much
weaker than in the case of constant EF.

\begin{figure}
  \centering
  \includegraphics[width=0.32\textwidth]{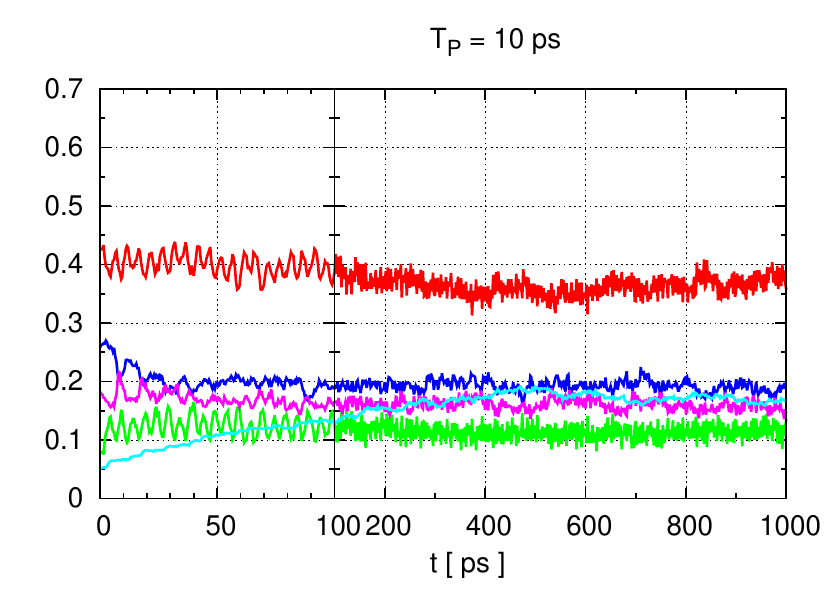}
  \includegraphics[width=0.32\textwidth]{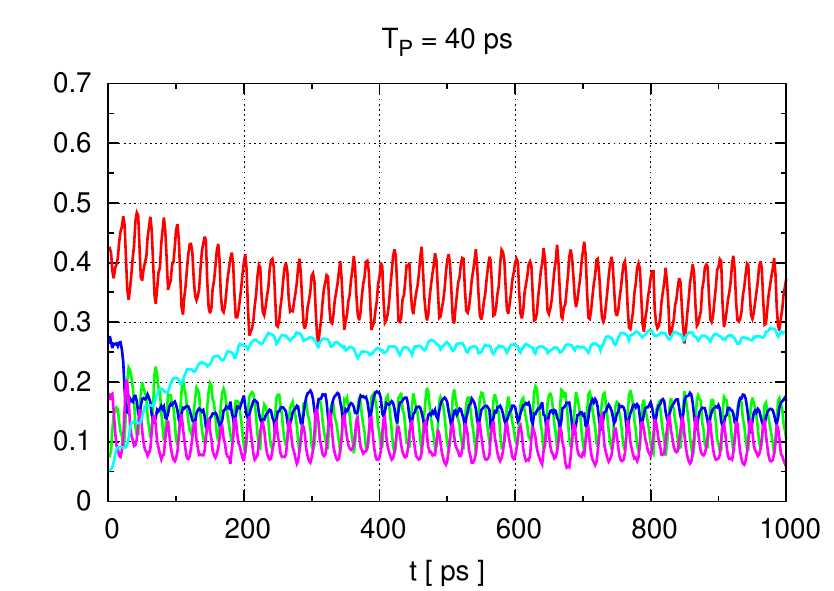}
  \includegraphics[width=0.32\textwidth]{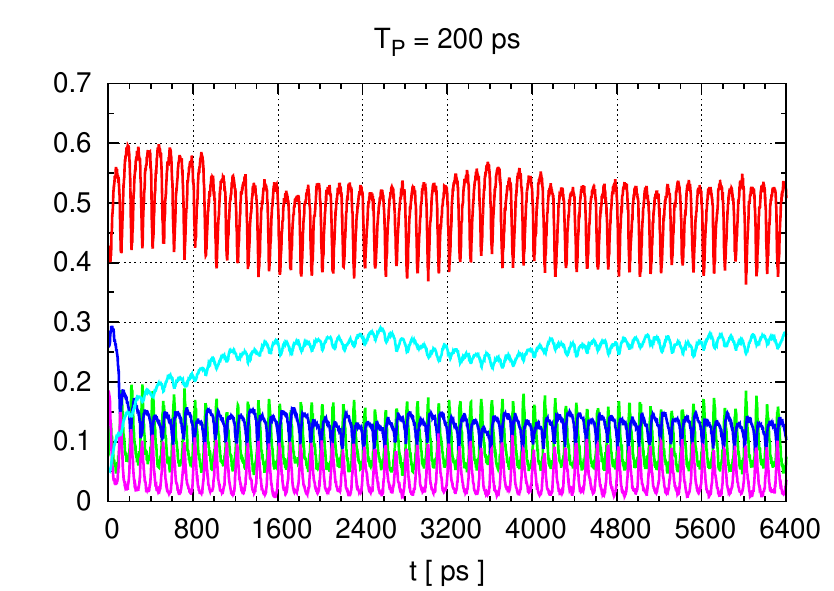}
  \caption{The probability of being in a given of conformation for alanine in a   periodically oscillatory EF. Different periods,
    i.e. $\period =10$~ps, 40~ps and 200~ps, are considered here.  The
    probability of being in a certain conformation $I$, which is
    denoted by $P_I$ in this paper, is plotted by colored lines against
    time. Red line: $I = \confaa$,  green line: $I = \confab$,  dark
    blue line: $I = \confba$,  pink line: $I = \confbb$ and  light blue
    line: $I = \confc$. For $\period =10$~ps and 40~ps, the nonequilibrium
    simulations last 1000~ps, while for $\period =200$~ps,
    they last 6400~ps.  }
  \label{fig:tmp9}
\end{figure}

\begin{figure}
  \centering
  \includegraphics[width=0.19\textwidth]{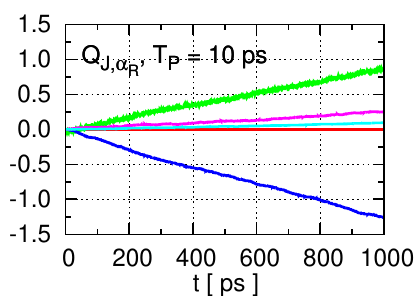}
  \includegraphics[width=0.19\textwidth]{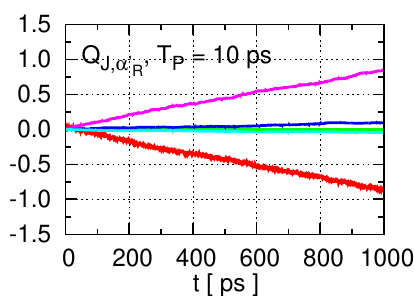}
  \includegraphics[width=0.19\textwidth]{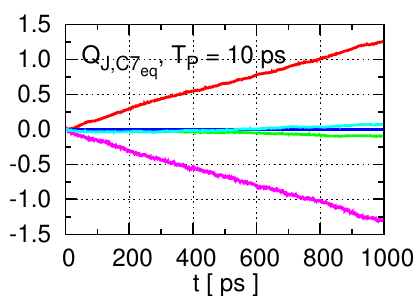}
  \includegraphics[width=0.19\textwidth]{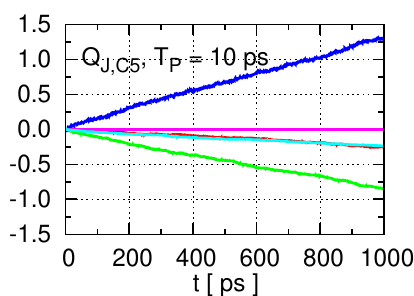}
  \includegraphics[width=0.19\textwidth]{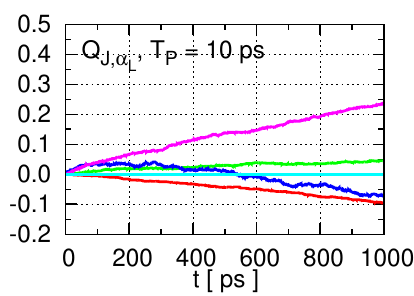}\\
  \includegraphics[width=0.19\textwidth]{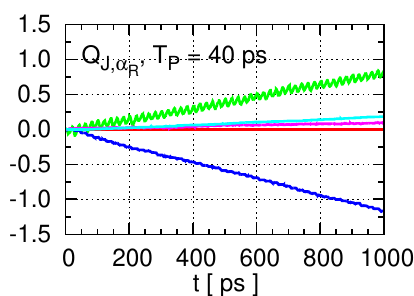}
  \includegraphics[width=0.19\textwidth]{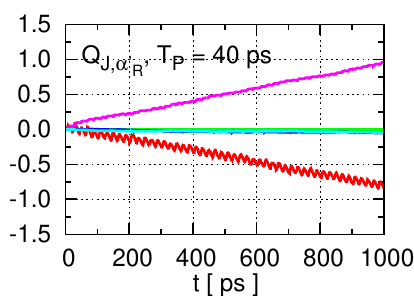}
  \includegraphics[width=0.19\textwidth]{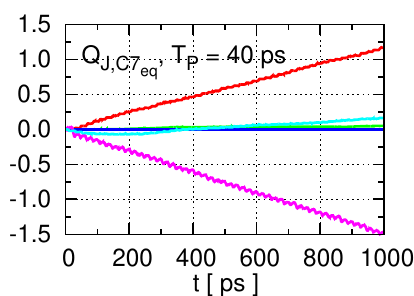}
  \includegraphics[width=0.19\textwidth]{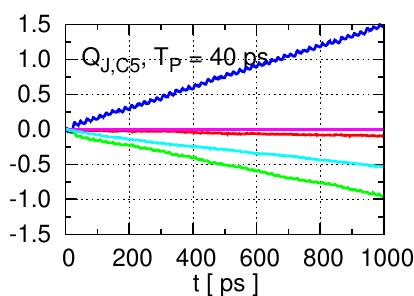}
  \includegraphics[width=0.19\textwidth]{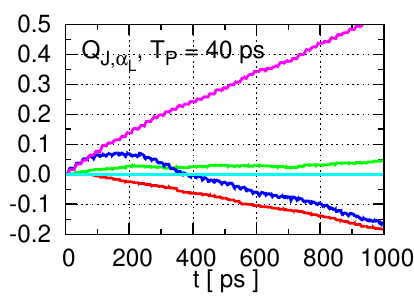}\\
  \includegraphics[width=0.19\textwidth]{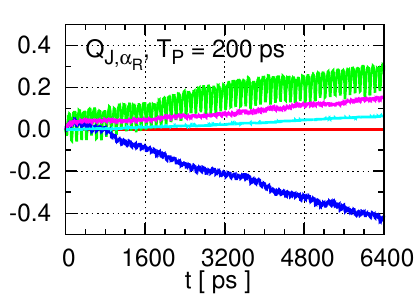}
  \includegraphics[width=0.19\textwidth]{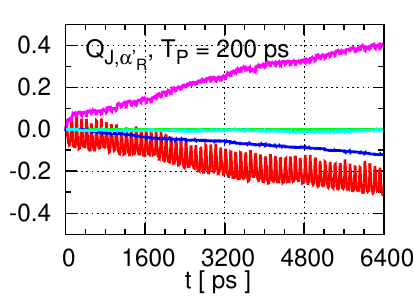}
  \includegraphics[width=0.19\textwidth]{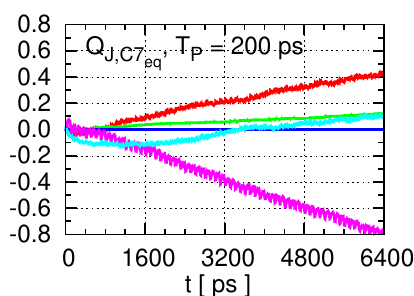}
  \includegraphics[width=0.19\textwidth]{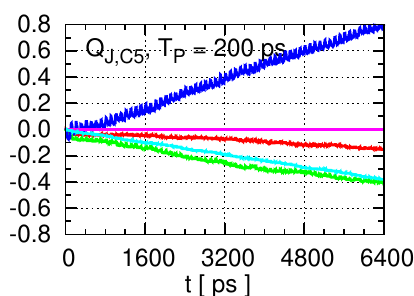}
  \includegraphics[width=0.19\textwidth]{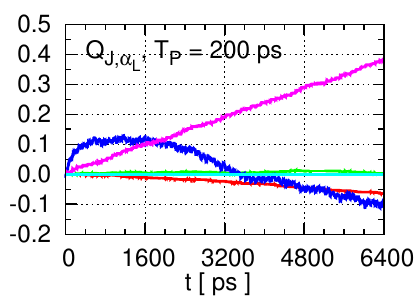}\\
  \caption{
    Integrated probability flux $Q_{J,I}(t)$ for the periodically
    oscillatory EF, is plotted against time (horizontal axis, in picoseconds).
    The integrated
    probability flux is defined by $Q_{J,I}(t) = \int_0^t F_{J,I}(\tau) d\tau$, where
    $F_{J,I}(t)$ is the net probability flux from conformation $J$ to $I$.
    See also Eqn.~\eqref{eqn:tmp7} and~\eqref{eqn:tmp7a} for definitions
    of $F_{J,I}(t)$ and $Q_{J,I}(t)$, respectively.
    From top to bottom, rows report results with  period 10, 40 and 200~ps, respectively.
    From left to right, the five
    columns show the integrated flux $Q_{J,\confaa}$, $Q_{J,\confab}$,
    $Q_{J,\confba}$, $Q_{J,\confbb}$ and $Q_{J,\confc}$, respectively. In each plot,
    the red line stands for $J=\confaa$, green for $J=\confab$, dark blue for $J=\confba$,
    purple for $J=\confbb$ and light blue for $J=\confc$.
    An increasing value of $Q_{J,I}$ indicates a net probability
      flux from conformation $J$ to $I$,
      while a decreasing value of $Q_{J,I}$ indicates a net probability
      flux from $I$ to $J$.
  }
  \label{fig:tmp10}
\end{figure}

Fig.~\ref{fig:tmp9} shows the probability of the
conformations against time for periodically oscillatory EF,
while  
Fig.~\ref{fig:tmp10} presents the integrated net probability fluxes
(defined by Eqn.~\eqref{eqn:tmp7a}) between
conformations. Here the probability fluxes are not reported
because the profiles are highly oscillating and they would not offer 
a better information than the one that can be obtained from the integrated probability flux.
Coming back to Fig.~\ref{fig:tmp9}, for all periods investigated,
the observed time-dependent probabilities  in $\confaa$, $\confab$, $\confba$ and $\confbb$ are
highly oscillating and the average value over time cycles does not change
considerably w.r.t.~time.
However, the probability in conformation $\confc$ significantly
increases to approximately 0.17 for $\period =10$~ps, 0.27 for  $\period =40$~ps, and
0.26 for $\period =200$~ps.
In the case of $\period =10$ and $40$~ps, {the probability of conformation $\confc$}
reaches the steady value in around 300~ps, while it requires about 1200~ps
for the $\period =200$~ps case. 
The notable increment of probability of $\confaa$ and
  the vanishing of $\confba$ in the constant EF case (see Fig.~\ref{fig:tmp5})
are not observed in the oscillatory EF case.
\begin{figure}
  \centering
  \includegraphics[width=0.6\textwidth]{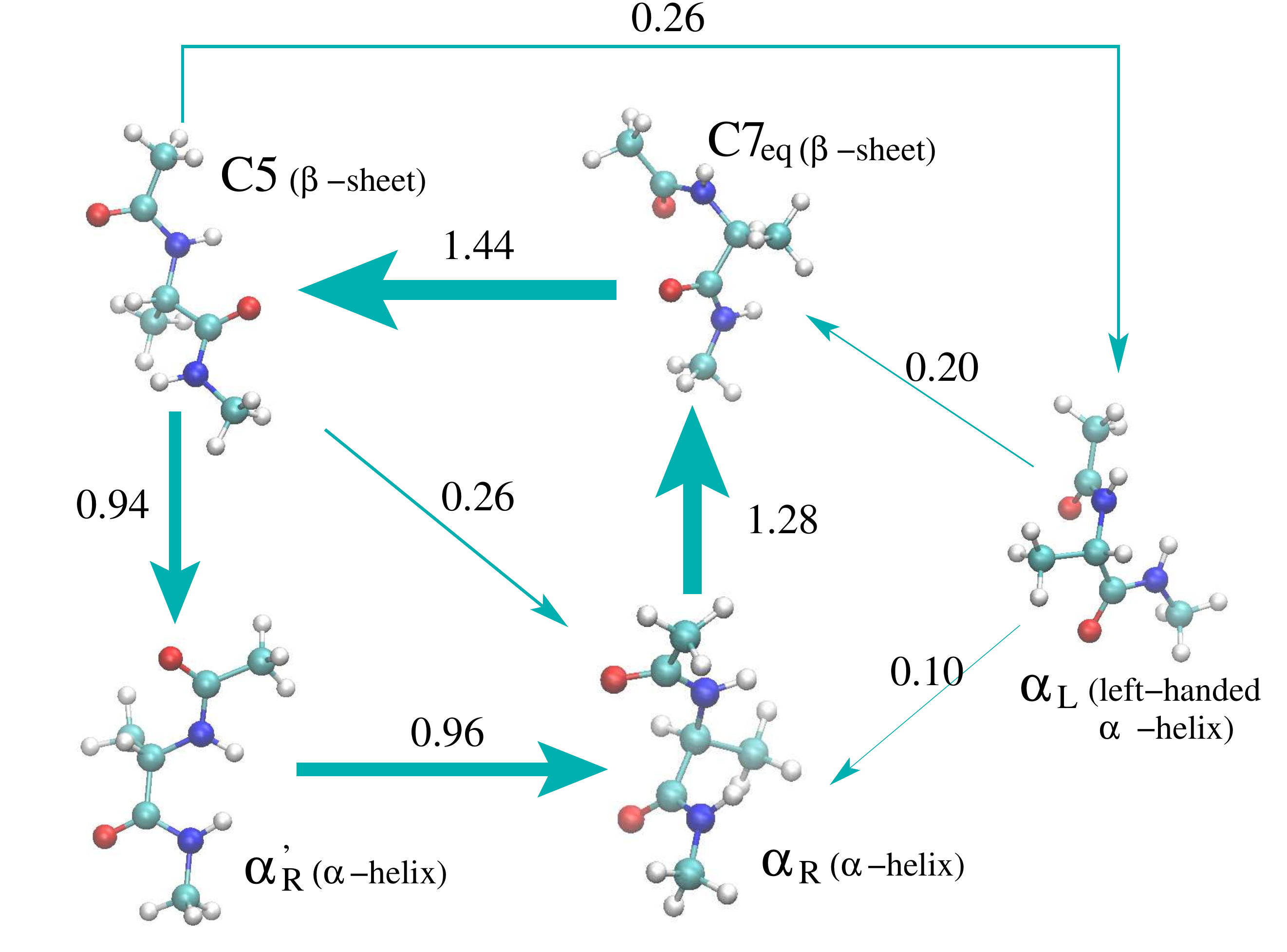}
  \caption{Schematic plot of the main probability fluxes between conformations
    for oscillatory EF, $\period =10$~ps. The thickness 
    of the arrows approximately presents the strength of the flux.
    The numbers near the arrows indicates the strength of the averaged
    probability flux, the unit of which is $10^{-3}\times\textrm{ps}^{-1}$.
  }
  \label{fig:tmp11}
\end{figure}
In comparison to the static EF case, where the probability flux
vanishes when the system reaches steady state (Fig.~\ref{fig:tmp6}), 
Fig.~\ref{fig:tmp10} shows increasing
integrated probability net fluxes, which imply some lasting and directional
fluxes among the conformations {in steady state}.
Fig.~\ref{fig:tmp11} presents
a schematic plot of the main probability fluxes for $\period =10$~ps.
The thickness of the arrows and the numbers nearby indicate
the steady value of average net fluxes over periods.
The relevance of Fig.~\ref{fig:tmp11} is in the fact that it suggests,
in perspective, a general scenario common to all molecules
characterized by $\beta$-sheet-like or $\alpha$-helix-like
conformations and how an electric field is likely to modify it.
  Although the network presented in Fig.~\ref{fig:tmp11} is similar to those
  discovered in the equilibrium studies 
  in the sense of the steady conformations and connections among them (e.g.~in Ref.~\cite{apostolakis1999calculation, gfeller2007complex}),
  the physical meaning is quite different in our study:
  The system is under nonequilibrium conditions due to a controlled physical external perturbation, and
  the related net probability fluxes are based on trajectories generated by such a perturbation. In contrast, in the equilibrium studies,
  such net probability fluxes do not explicitly exist.
The probability flux of $\period =40$~ps is quantitatively
comparable to the case $\period =10$~ps, except that the probability fluxes
going into conformation $\confc$ is stronger, which actually results in
a higher steady probability in $\confc$.
The probability flux of case $\period =200$~ps is qualitatively
similar to $\period =10$~ps and $\period =40$~ps, however,
the strength is much lower than the latter two cases.
Moreover, 
  we see that the integrated fluxes reach
  a steadily
increasing stage after about 3200~ps, which is even longer than the time scale
at which the probability of $\confc$ reaches its steady value. This indicates
a longer intrinsic time scale in the $\period =200$~ps case.

We have not studied periods longer than $\period =200$~ps, because
the trend suggests that a longer period indicates even longer intrinsic time scales.
However, the long-period-limit case can be safely guessed: When
the EF changes so slowly that
at each time the system
can be viewed as in equilibrium,
the process falls in the category of a quasi-equilibrium
process. 

\section{Discussion and Conclusion}
We have investigated the conformational
dynamics of a solvated alanine dipeptide under the action of a constant and oscillatory electric
field (EF).
We have employed the dynamical nonequilibrium molecular dynamics
(D-NEMD) method. This allowed us to analyze the conformational changes
of the molecule in terms of a response to an external perturbation
which drives the system away from equilibrium. From the technical
point of view, we have proposed a local thermostating procedure which
does not introduce invalidating artifacts and at the same time avoids
the necessity of considering large water bulk systems to solvate the
molecule.
The conformations of the alanine dipeptide are
firstly projected onto the Ramachandran plot, and
then grouped in 5  conformations.
These conformations
correspond to different secondary structures in larger molecules
(i.e.~proteins).
Next, the time-dependent probability
of being in a certain conformation and the net probability flux
between them are calculated.
We have compared the case of constant EF and that of oscillatory EF
and reported the main differences.
Worth to notice is the possibility of employing an EF in order to identify several time scales related to conformational changes, this would not be straightforward for standard equilibrium MD simulations and thus suggest that D-NEMD proposed here is a powerful tools to investigate the conformational properties of a large molecule in solution.

The intensity of the EF used in this paper is 1~V/nm for both constant
and oscillatory cases. This intensity is 4-5 orders of magnitude
higher than that reachable in a standard laboratory microwave
instrument~\cite{damm2012can}. However, this intensity is achievable
by a sharp electric field emitter tip~\cite{scovell2000phase}, near
the surface of mica~\cite{starzyk2013proteins}, or by a modern laser
equipment~\cite{vogel2008femtosecond}.
{The minimal frequency investigated here (5.00~GHz, i.e.~$\period = 200$~ps)
is of the same order of magnitude as the microwave
radiation available in laboratory (2.45~GHz)~\cite{damm2012can}.
The highest frequency (100~GHz, i.e.~$\period = 10$~ps) is
one order of magnitude lower than the tetrahertz spectroscopy experiments~\cite{plusquellic2007applications, born2009terahertz},
so our results cannot be directly connected to these experimental results.}
The authors are fully aware
that the EFs investigated in the present paper are, at this stage, still ideal, above all regarding their spatially homogeneity and
  the fact that the time-dependency is always well defined (as pointed
out by previous simulation studies~\cite{budi2005electric,
  budi2007effect, budi2008comparative, toschi2008effects,
  astrakas2011electric, astrakas2012structural, damm2012can,
  starzyk2013proteins, english2009nonequilibrium,
  solomentsev2012effects}). However, the present work is an attempt to understand how the use of D-NEMD as a simulation tool can be applied to the important case of an EF and which information one may extract that is not easily accessible to other MD procedures (e.g.  EF induced conformational changes and its related time scales which opens the
possibility of conformational manipulation of the protein by applying
an external EF.)

In order to sharply define the advantages and limitations of the approach used we must also clarify that in order to describe the system, we use a classical force field with TIP3P water model.
The EF is well know to be able to protonate the water molecule at the
intensity we used in the paper. However, this happens at the time
scale of femtoseconds, which is much shorter than the slowest time
scale resolved by the current research, so on average it should not play an essential role on the conformational dynamics.
The polarization w.r.t.~the external EF also does
not play a important role, because  under the EF of 1~V/nm, the induced dipole
of water is only c.a.~0.05~Debye that is negligible comparing with the
dipole moment response observed in our simulations.  Therefore, it is
reasonable to use a non-polarizable model in this study.  The reason we
prefer a rigid water model is that it is computationally more
efficient than any polarizable water model.  We also want to remind
the reader that it has been reported recently that the physical scale of the slowest
dynamics differs in a considerable way between different force
fields~\cite{vitalini2013speed}. However, the main qualitative conclusions, for
example those concerning the determination of time scales or the map of conformational changes, 
are not likely to change significantly as a function of the force field.

\section*{Acknowledgment}
  H.W. acknowledges the valuable discussions with Simone Meloni and Burkhard Schmidt.
  This work was partially supported by the Deutsche
  Forschungsgemeinschaft (DFG), with the Heisenberg
  Grant provided to L.D.S. (Grant No. DE 1140/5-1),
  and with the support within the priority program ``Ionic Liquids'' SPP 1191
  (H.W. and L.D.S.). 
  H.W.~and C.S.~thank the financial support by DFG research center MATHEON.
  Funding from the IIT SEED project SIMBEDD and from the SFI Grant 08-IN.1-I1869 is acknowledged by G.C..

  \appendix
\section{Comparison with the linear response theory}  
\begin{figure}
  \centering
  \includegraphics{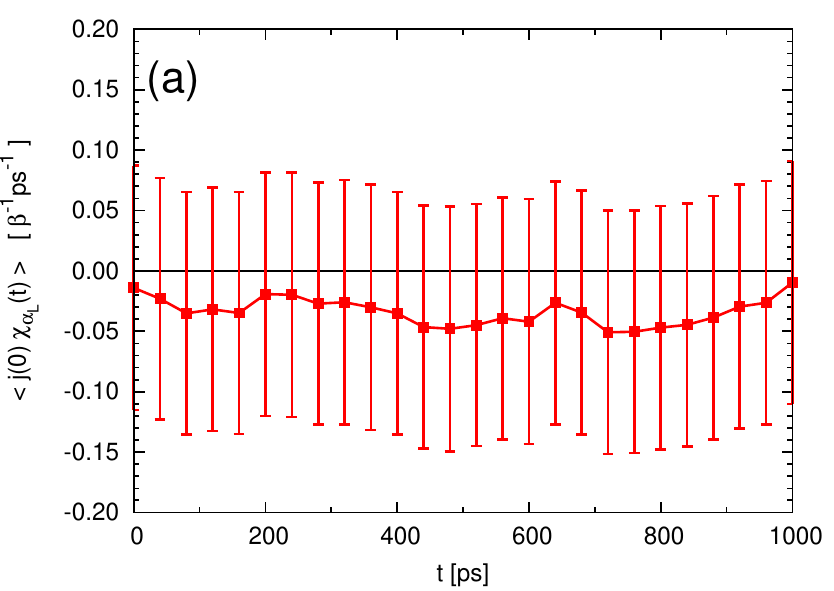}
  \includegraphics{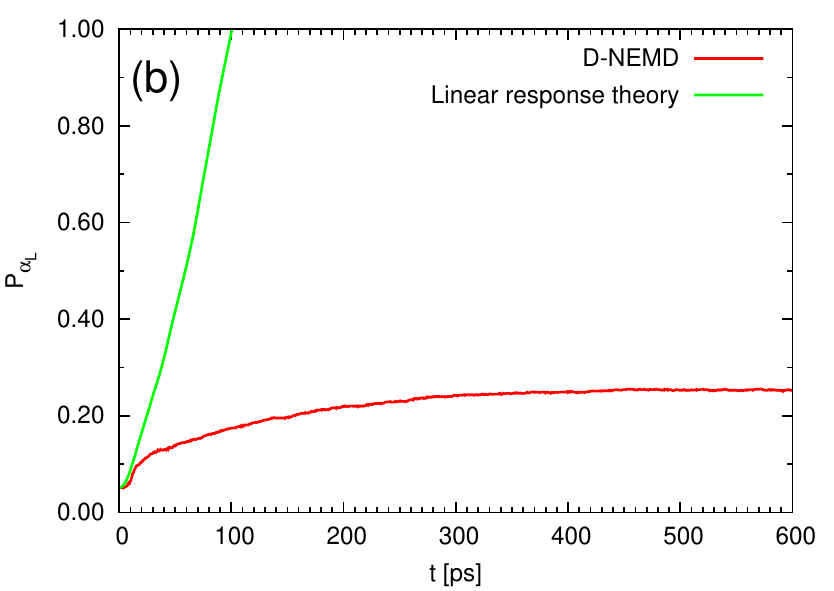}
  \caption{
    {The numerical result of the linear response theory.
    (a): The correlation function $\langle j(0)\cdot \chi_{\confc}(t)
    \rangle_0$ against time $t$.
    The error bars in the figure denote
    the statistical uncertainty at 95\% confidence level. 
    (b): The probability of conformation $\confc$ computed from the
    linear response theory~\eqref{eqn:appdx1} versus that
    computed from D-NEMD~\eqref{eqn:tmp1}.}
  }
  \label{fig:appdx1}
\end{figure}

A commonly used equilibrium approach that can describe the nonequilibrium
response is the linear response theory (Green-Kubo relation~\cite{green1954markoff,kubo1957statistical}).
We employed such a method to test the
conformational dynamics of alanine dipeptide under the EF.
The response function, for example the time-dependent probability of
left-handed $\alpha$-helix, is defined as~\cite{tuckeman2010statistical}
\begin{align}\label{eqn:appdx1}
  P_\confc(t) = \langle\chi_{\confc}(t)\rangle = \langle \chi_{\confc} \rangle_0 -
  \beta \int_0^t ds\; M_e(t - s)\langle j(0)\cdot \chi_{\confc}(t) \rangle_0,
\end{align}
where $\chi_{\confc}$ is the characteristic function of set $\confc$ that takes
the value of 1 for $(\phi,\psi)\in \confc$, and takes value 0
otherwise. $M_e(t)$ is the relative magnitude of the EF. For the case of
constant EF, for example, $M_e(t) = t/t_{warm}$ for $0\leq t<t_{warm}$, and
$M_e(t) = 1$ for $t\geq t_{warm}$.
$j$ is the dissipative flux that is defined by
\begin{align}
  j = - \sum_{i=1}^N E_\infty q_i v_{i,x},
\end{align}
where $q_i$ is the partial charge of the $i$th atom, and $v_{i,x}$ is
the $x$ component of the velocity of the $i$th atom.  The notation
$\langle\cdot\rangle_0$ denotes the equilibrium ensemble average.
We report the correlation function
$\langle j(0)\cdot \chi_{\confc}(t) \rangle_0$ as a function of correlation time $t$
in Fig.~\ref{fig:appdx1}~(a). The value of the function is estimated from
two independent 1~$\mu$s equilibrium simulations.
It must be noticed that the total length of the equilibrium
trajectories is the same as the nonequilibrium simulation (2000
trajectories, each 1000~ps long).
Results show that the statistical error is much larger than the value of
correlation function itself.
  In Fig.~\ref{fig:appdx1}~(b),
  the probability of conformation $\confc$ calculated from
  the linear response theory~\eqref{eqn:appdx1} is compared with that
  computed from D-NEMD~\eqref{eqn:tmp1}.
  The linear response result follows the D-NEMD result only in the very first
  15~ps, then it diverges. Since it is meaningless for a probability
  being larger than 1, the linear response result is qualitatively wrong.
Therefore, the
overwhelming statistical uncertainty suggest to prefer
D-NEMD~Eq.~\eqref{eqn:tmp1} to calculate the
nonequilibrium averages. 
In fact, since the total computational effort of
the equilibrium simulation is the same as the nonequilibrium
simulation, the nonequilibrium simulation is more
accurate than the conventional equilibrium approach (linear response) at the same computational cost.
A similar
observation has already been reported in Ref.~\cite{ciccotti1975direct}.

\end{document}